\documentstyle[floats,aps,psfig,pre]{revtex}
\begin{document}
\flushbottom
\twocolumn[\hsize\textwidth\columnwidth\hsize\csname @twocolumnfalse\endcsname

\title{
Self-avoiding walks and connective constants in small-world networks}
\author{Carlos P. Herrero$^1$ and Martha Saboy\'a$^2$}
\address{
$^1$Instituto de Ciencia de Materiales,
    Consejo Superior de Investigaciones Cient\'{\i}ficas (CSIC),
    Campus de Cantoblanco, 28049 Madrid, Spain \\
$^2$ Facultad de Ciencias Econ\'omicas y Empresariales,
    Universidad Aut\'onoma de Madrid, 28049 Madrid, Spain } 
\date{\today}
\maketitle

\begin{abstract}
 Long-distance characteristics of small-world networks have been
studied by means of self-avoiding walks (SAW's).  We consider networks 
generated by rewiring links in one- and two-dimensional regular lattices.
The number of SAW's $u_n$ was obtained from numerical simulations as a 
function of the number of steps $n$ on the considered networks. 
The so-called connective constant, $\mu = \lim_{n \to \infty} u_n/u_{n-1}$, 
which characterizes the long-distance behavior of the walks, increases 
continuously with disorder strength (or rewiring probability, $p$).
For small $p$, one has a linear relation $\mu = \mu_0 + a p$,
$\mu_0$ and $a$ being constants dependent on the underlying lattice.
Close to $p = 1$ one finds the behavior expected for random graphs. 
An analytical approach is given to account for the results derived from 
numerical simulations.  Both methods yield results agreeing with each other 
for small $p$, and differ for $p$ close to 1, because of the different 
connectivity distributions resulting in both cases.
\end{abstract}


\vskip2pc]
\narrowtext

\section{Introduction}

Our world is formed by networks of different types (social, biological, 
technological, economic), whose characterization has launched in last
years the emergence of models incorporating the basic ingredients 
of real-life networks \cite{st01,al02,do02}.
In particular, social networks form the substrate where processes such
as information spreading or disease propagation take place. One expects
that the structure of these complex networks will play an important role
in such dynamical processes, which are usually studied by means of stochastic
dynamics and random walks. Some processes, such as navigation and
exploratory behavior are neither purely random nor totally deterministic, 
and can be also described by walks on graphs \cite{vi99}.  
In this context,
the generic properties of deterministic navigation \cite{li01} and directed
self-avoiding walks \cite{sa01} in random networks have been analyzed recently.

In last years, networks displaying the ``small-world'' effect
have been intensively studied \cite{wa98,la00,la01a,ne00,al99}. 
Watts and Strogatz\cite{wa98,wa99} proposed for this kind of networks a model
based on a locally connected regular lattice,
in which a fraction $p$ of the links between nearest-neighbor sites
are replaced by new random connections, thus creating long-range ``shortcuts''. 
Hence, one has in the same network a local neighborhood (as for regular 
lattices) and some global properties of random graphs \cite{bo85}.
The small-world effect is usually measured
by the scaling behavior of the characteristic path length $\ell$, 
defined as the average of the distance between any two sites.
In small-world networks, one has a logarithmic increase of $\ell$
with the network size, as happens for random graphs \cite{al02,bo85,ca00}.

 This short global length scale changes strongly the behavior
of statistical physical problems on small-world networks, as compared
with regular lattices (where one has $\ell \sim N^{1/d}$, $N$ being the
system size and $d$ the lattice dimension).
Among these problems, one finds signal propagation \cite{wa98}, 
spread of infections \cite{ku01,mo00}, 
and random spreading of information \cite{pa01,la01b,he02b}.
Site and bond percolation \cite{mo00,ne99}, as well as
the Ising \cite{ba00,gi00,he02a} and XY models \cite{ki01}, have been also studied
in these networks.
Most of the published work on small worlds has focussed 
on networks obtained from one-dimensional lattices (rings).
Small-world networks built by rewiring lattices of higher dimensions
have being employed to study percolation, as a model of disease propagation
\cite{ne99,ne02}.
Several characteristics of random walks on this kind of networks have been
analyzed in connection with diffusion processes \cite{je00a,ja01}. In particular,
some properties of these walks, such as the probability of returning to the origin,
were found to be intermediate between those corresponding to fractals and Cayley 
trees \cite{je00b}.

In this paper we study self-avoiding walks in small-world networks 
built up from one- and two-dimensional regular lattices.
A self-avoiding walk (SAW) is defined as a walk along the bonds of
a given network which can never intersect itself. The walk is
restricted to moving to a nearest-neighbor site during each step,
and the self-avoiding condition constrains the walk to occupy only
sites which have not been previously visited in the same walk \cite{do69}.
SAW's have been used for modeling the large-scale properties of
long-flexible macromolecules in solution \cite{ge79}, as well as
for the study of polymers trapped in porous media, gel
electrophoresis, and size exclusion chromatography, which deal with the
transport of polymers through membranes with small pores \cite{le89}. 
They have been also employed to characterize complex crystal structures  
\cite{he95} and to analyze critical phenomena in lattice models \cite{do69,kr81}.  
Universal constants for SAW's have been discussed by Privman {\em et al.} 
\cite{pr91}.

The paper is organized as follows. 
In Sec.\,II we give some basic definitions and concepts related to SAW's.
In Sec.\,III we present results for SAW's on simulated small-world networks,
and in Sec.\,IV we give an approximate analytical procedure to calculate the
number of SAW's on this kind of networks.
The paper closes with some conclusions in Sec.\,V.

\section{Basic definitions}
 For regular lattices, the number $u_n$ of different SAW's starting
from a generic site has an asymptotic dependence for large $n$:
\cite{pr91,mc76}
\begin{equation}
         u_n \sim n^{\gamma - 1}   \mu^n  \hspace{2mm} ,
\label{un}
\end{equation}
where $\gamma$ is a critical exponent which depends on the lattice dimension,
and $\mu$ is the so-called ``connective constant'' or effective coordination
number of the considered lattice \cite{mc76,ra85}.
 In general, for a lattice with coordination number (or connectivity) $z$, 
one has $\mu \le z - 1$.
 This parameter $\mu$ can be obtained from Eq.\,(\ref{un}) by the limit:
\begin{equation}
       \mu = \lim_{n\to\infty} \frac{u_n}{u_{n-1}}
        \hspace{3mm} .
\label{mu}
\end{equation}            
The connective constant depends upon the particular topology of each 
lattice, and has been determined very accurately for two-dimensional (2D) 
and three-dimensional (3D) lattices \cite{so95}. 
In the following, we will consider Eq. (\ref{mu}) as a definition of the
connective constant $\mu$ for any network. [Note that the limit in 
Eq. (\ref{mu}) is well defined provided that the mean connectivity is
finite.]

 For random and small-world networks the number of SAW's of length $n$
depends on the considered starting node of the network.
In the sequel we will call $u_n$ the average number of SAW's of length $n$,
i.e. the mean value obtained (for each $n$) by averaging over the 
network sites.
For small-world networks one expects $\mu$ values larger than
that corresponding to the starting regular lattice. In particular,
$\mu$ is expected to increase with $p$ and approach the value
corresponding to random lattices with mean connectivity $z$
as $p \to 1$.     

For regular lattices all nodes have the same connectivity,
i.e., the same number of nearest neighbors. However, for
$p > 0$ different connectivities $m$ are possible, giving rise to a
probability distribution for which analytic expressions have been
found \cite{he02b,ba00}.
For a (large) random network with
mean connectivity $z$, the connectivity distribution $P_{rd}(m)$
follows a Poisson law \cite{al02,bo85}:
\begin{equation}
P_{rd}(m) =  \frac{z^m \, e^{-z}}{m!}  \, .
\label{rd}
\end{equation}

For random networks, the connective constant can be obtained in a
straightforward manner, due to the absence of correlations between links. 
For $n = 1$ one has obviously $u_1^{rd} = z$.
Now, given a generic node and a link going out from it, the connectivity
distribution $Q_{rd}(m)$ for the other end of the link in a random graph
is given by
\begin{equation}
Q_{rd}(m) = \frac{m}{z} \; P_{rd}(m) \;  .
\label{qrd}
\end{equation}          
Then, the average number of two-step SAW's is given by
\begin{equation}
 u_2^{rd} = z \sum_{m>1} (m-1) Q_{rd}(m) \;  ,
\label{crd}
\end{equation}
and we find $u_2^{rd} = z^2$.
Using the same procedure for $n > 2$, one has:
\begin{equation}
   u_n^{rd} = z^n  \; ,
\label{unrd}
\end{equation}
and thus for (large) random networks one finds $\mu = z$.
In connection with this, we note that for a Bethe lattice (or Cayley
tree) with connectivity $z$, the number of SAW's is given
by  $u_n^{BL} = z (z-1)^{n-1}$, and one has $\mu = z - 1$ 
(see, e.g. Ref. \cite{st92}).

\section{Numerical simulations}
The networks studied here have been generated from three different
regular lattices: one-dimensional (1D) ring with coordination number 
$z$ = 4 and 6, and 2D square lattice ($z = 4$).  
To construct our small-world networks, we consider 
in turn each of the bonds in the starting lattice and substitute it with a 
given probability $p$ by a new bond. This means that one end of the bond 
(chosen at random) is changed to a new node taken randomly from the entire 
network.  We impose three conditions: 1) no two nodes can have more than
one bond connecting them, 2) no node can be connected by a link
to itself, and 3) isolated sites (with zero links) are not allowed.
This method keeps constant the total number of links in the
rewired networks (and consequently the average connectivity $z$). 
The total number of rewired links is $\frac{1}{2} z p N$ on average.

   For networks generated in the present way there is a 
crossover size $N^* \sim p^{-1}$ that separates the large- and
small-world regimes \cite{ba99b,me00}, and the small-world behavior 
appears for any finite value of $p$ ($0<p<1$) as soon as the network is 
large enough.  
 Our networks included $1 \times 10^5$ sites in 1D and
300 $\times$ 300 sites for the 2D system, so that we were in the 
small-world regime (system size $N > N^*$). In the sequel we will call $L$
the side length of the considered lattices, i.e. $L = N^{1/d}$.
Periodic boundary conditions were assumed. 
We note that our networks differ from those discussed by 
Watts and Strogatz \cite{wa98} in that these authors left untouched $z/2$ 
links per site.
 For our simulated small-world networks we have obtained the average number 
$u_n$ of SAW's up to $n$ = 21. Since $u_n$ increases with $p$, this maximum 
$n$ was reduced to $n = 14$ close to $p=1$, in order to carry out averages 
over nodes of the generated networks.  These numbers of steps in the SAW's
are sufficient to obtain the connective constant $\mu$ with enough accuracy
for our present purposes. In fact, the larger is $p$, the faster the ratio 
$u_n/u_{n-1}$ converges with $n$.

First of all, we calculate the mean-squared end-to-end distance of the walks
on our small-world networks:
\begin{equation}
R_n^2 = \langle ({\bf r}_n - {\bf r}_0)^2 \rangle  \;  ,
\label{rn}
\end{equation}
where $\langle \, \rangle$ indicates an average over $n$-steps SAW's with 
different starting sites and for different network realizations with
given $p$. Here, ${\bf r}_n$ refers to the position of site $n$ in the 
$d$-dimensional Euclidean space of the underlying regular lattice,
and ${\bf r}_0$ is the position of the origin for the considered walk. 
For SAW's on the 1D lattices considered here, one has $R_n^2 \sim b n^2$, whereas
for the 2D square lattice $R_n^2 \sim b n^{3/2}$ \cite{do69,mc76}, 
with a lattice-dependent constant $b$ of order unity in all cases.

\begin{figure}
\vspace*{-1.7cm}
\centerline{\psfig{figure=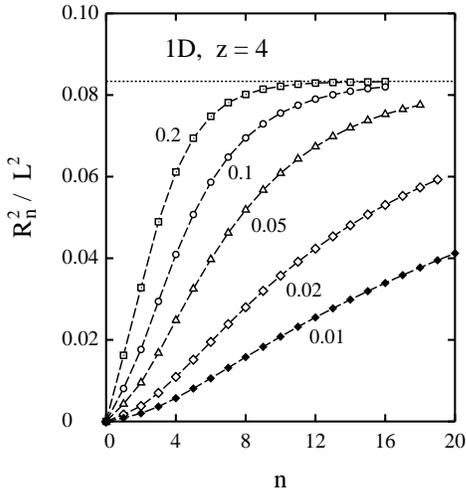,height=10.0cm}}
\vspace*{-1.5cm}
\caption{
Mean-square end-to-end distance $R_n^2$ for SAW's on small-world networks
generated from a 1D regular lattice with $N = 10^5$ sites and connectivity
$z = 4$.  We present values for $R_n^2$, normalized by the squared system
length $L^2$, as a function of the number of steps $n$. Different
symbols correspond to several values of the rewiring probability $p$.
From top to bottom: $p$ = 0.2, 0.1, 0.05, 0.02, and 0.01.
A dotted line indicates the value $R_n^2 / L^2 = 1/12$ corresponding to
$p = 1$. Dashed lines are guides to the eye.
} \label{f1} \end{figure}

In Fig. \ref{f1} we present the ratio $R_n^2/L^2$ as a function of the number 
of steps $n$ for SAW's on 1D small-world networks with $z = 4$, for several
values of the rewiring probability $p$. $R_n^2$ increases with
$n$ much faster than for the corresponding regular lattice 
due to the random connections introduced by the
rewiring of links. (Note that for a regular lattice with $L=10^5$, we have 
for $n=20$ a ratio $R_n^2/L^2 \sim 10^{-7}$.) 
In fact, if we call $f_n$ the fraction of $n$-steps
paths that include at least one rewired link, we have:
\begin{equation}
R_n^2 \approx b (1 - f_n) n^2 + \frac{1}{12} f_n d L^2  \;  .
\label{rn2}
\end{equation}
The first and second terms on the r.h.s. come from SAW's without and 
with rewired links, respectively. The second term amounts, apart from
the fraction $f_n$, to the average Euclidean distance between any pair of 
sites in a $d$-dimensional box with side length $L$. 
Thus, in the limit of large networks (as those 
considered here, with $L \gg n$) one has  $R_n^2 \approx f_n d L^2 / 12$.
In the course of our numerical simulations we have checked Eq.\,(\ref{rn2}) 
by calculating independenlty $f_n$ and $R_n$ as a function of $n$. 
We found that both sides of this equation coincide within error bars 
(which are smaller than the symbol size in Fig. \ref{f1}).  Then, $R_n^2$ can be
considered as a measure of the fraction of SAW's containing rewired links,
and converges to $d L^2/12$ for large $n$ in the rewired networks ($f_n \to 1$).

\begin{figure}
\vspace*{-1.7cm}
\centerline{\psfig{figure=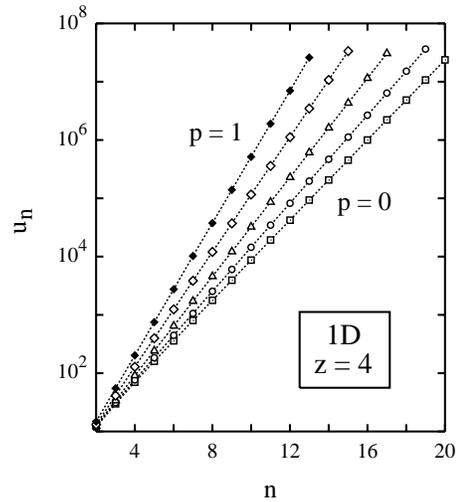,height=10.0cm}}
\vspace*{-1.5cm}
\caption{
Average number of self-avoiding walks $u_n$ on small-world networks generated from 
1D rings with connectivity $z=4$. We plot $u_n$ as a function of the path length
$n$ for several rewiring probabilities $p$, as derived from numerical simulations.
From top to bottom: $p$ = 1, 0.3, 0.1, 0.03, and 0.
Dotted lines are guides to the eye.
} \label{f2} \end{figure} 

We now turn to the number $u_n$ of SAW's on these networks. 
In Fig. \ref{f2} we show $u_n$ as a function of the walk length for networks
built up from a 1D lattice with coordination number $z=4$. We have plotted
results for several rewiring probabilities $p$, from $p=0$ (regular lattice,
squares) to $p=1$ (black diamonds). As expected, $u_n$ increases as $p$ is 
raised, since introducing long-range connections in the starting lattice 
opens new ways for the SAW's. In particular, such long-range links allow the
walks to visit regions far away from the origin for small $n$, and thus avoid 
the constriction
associated to move close to the starting site, which limits the number of
possible self-avoiding walks.
In the logarithmic plot of Fig. \ref{f2} one sees that $\log(u_n/u_{n-1})$ 
converges rather fast to a constant for each rewiring probability $p$, which
allows us to calculate the corresponding connective constant.
Also, from the results shown in Fig. \ref{f2} we find that for large $n$ the
ratio between the number of SAW's for $p=1$ and $p=0$ increases as
$k^n$ with a constant $k$ = 1.67. This means that at long distances, for each
link available for SAW's in the regular lattice, we have in average 1.67 connections
for $p=1$. This number is in fact the ratio between connective constants
for $p=1$ and $p=0$ in the case $d=1$, $z=4$.

\begin{figure}
\vspace*{-1.7cm}
\centerline{\psfig{figure=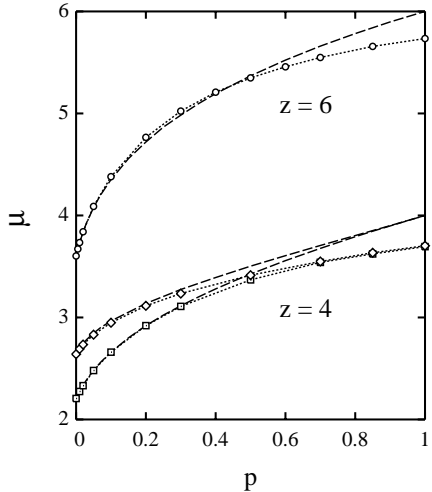,height=10.0cm}}
\vspace*{-1.5cm}
\caption{
Connective constant $\mu$ as a function of the rewiring probability
$p$. Different symbols represent results obtained for small-world networks
generated from 1D regular lattices with connectivity $z$ = 4 (squares)
and $z$ = 6 (circles), as well as from a 2D square lattice (diamonds).
Dotted lines are guides to the eye.  Error bars are less than the symbol size.
Results for $\mu$ obtained by means of the analytical method described in
Sec. IV are plotted as dashed lines.
} \label{f3} \end{figure}

The connective constant $\mu$ has been obtained for our simulated networks
by finding the large-$n$ limit of the ratio $u_n/u_{n-1}$. 
In Fig. \ref{f3} we present the resulting $\mu$ as a function of 
the rewiring probability $p$ for our networks generated from 1D and 2D 
lattices.  One observes that $\mu$ changes fast close to $p = 0$, and
the derivative $d\mu/dp$ decreases as $p$ is raised.
The largest change of $\mu$ in the whole region between $p = 0$ and 1
is found for the networks with $z = 6$. In fact, we find in this case an increase 
in $\mu$ of 2.13 to be compared with 1.48 and 1.07 in 1D and 2D rewired networks 
with $z=4$.

 For $p = 1$ our numerical procedure gives in all cases connective
constants $\mu$ clearly lower than the mean connectivity $z$.  In fact,
we found $\mu$ = 3.69, 5.73, and 3.70 ($\pm 0.01$) for networks rewired from
1D lattices with $z$ = 4 and 6, and from a 2D lattice ($z = 4$), respectively.
The obtained values for networks with $z = 4$ coincide within error bars,
irrespective of the starting 1D or 2D lattice, indicating that for $p=1$
the resulting rewired networks lost memory of the starting regular lattice.
However, the $\mu$ values obtained for simulated networks with $p=1$
contrast with those expected for random networks, that coincide in each case
with the average connectivity $z$, as explained above.
This occurs because our networks with $p = 1$ are not true (poissonian) random
networks, since they still keep memory of the starting regular lattices \cite{he02b}.
This memory effect is due mainly to the fact that one rewires only one end of
each link, maintaining the other end on its original site. Hence, the connectivity
distribution found for our rewired networks with $p=1$ does not coincide with
that given above in Eq. (\ref{rd}).
Such a difference with poissonian random networks should be even stronger for 
small-world networks generated in a way similar to ours, but leaving untouched 
$z/2$ links per site, as those studied in earlier works \cite{wa98,ba00}.

\begin{figure}
\vspace*{-1.7cm}
\centerline{\psfig{figure=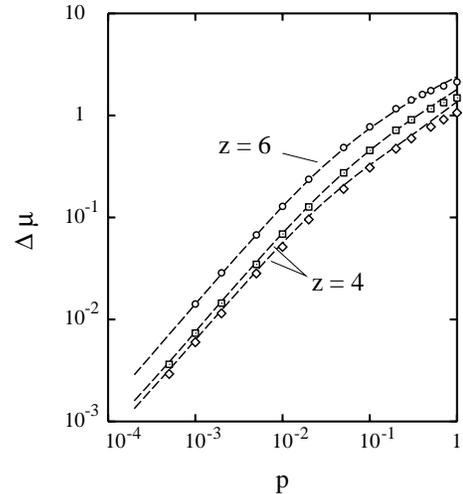,height=10.0cm}}
\vspace*{-1.5cm}
 \caption{
Dependence upon the rewiring probability $p$ of the change in connective
constant, $\Delta \mu = \mu - \mu_0$, with respect to the corresponding
regular lattices. Symbols represent the same networks as in Fig. \ref{f3}.
Dashed lines were obtained from analytical calculations by using
Eq. (\ref{un2}).
} \label{f4} \end{figure}

To analyze the change of $\mu$ as a function of $p$ for a given underlying
lattice, we call $\Delta \mu = \mu - \mu_0$, $\mu_0$ being the connective
constant of the corresponding regular lattice.
In Fig. \ref{f4} we show the obtained dependence of $\Delta \mu$ upon $p$
for the considered 1D and 2D networks in a log-log plot. In all three 
cases we find that $\Delta \mu$ can be fitted well by
a power law $\Delta \mu \sim p^c$ for $p \lesssim 0.01$.
For 1D networks, the exponent $c$ is found to be
$0.98 \pm 0.03$ (for $z = 4$) and $0.99 \pm 0.03$ (for $z = 6$).
For 2D networks we found $c = 0.99 \pm 0.03$.
Thus, our results indicate a linear dependence
$\mu = \mu_0 + a p $ for small $p$, irrespective of the underlying lattice.

Therefore, the functional form for the $p$-dependence of $\Delta \mu$
close to $p = 0$ does not depend on the dimension of the starting
regular lattice. This constrasts with other properties of
small-world networks, which have been found to change as
$p^{1/d}$, $d$ being the dimension of the underlying lattice. 
This happens for example for the average number of nodes in
``shell'' $n$, which scales as $p^{1/d}$ \cite{he02b}. Such a dependence is 
related to the scaling with $d$ of the characteristic length scale of small-world 
networks, namely the average distance between ends of shortcuts $\xi$,
given by $\xi = (p z)^{-1/d}$ \cite{ne99}.

\section{Analytical approximation}
We now derive an approximate analytical expression that allows us to 
calculate $u_n$ in a small-world network, assuming the sequence $\{u_n^0\}$ 
for the underlying regular lattice to be known.
For a given path length $n$ we obtain the mean number of SAW's $u_n$ by 
considering all possible sequences of unrewired and rewired links in small-world
networks.  With this purpose, we calculate the probability of reaching a 
rewired link as a function of the walk length.  
In particular, we will obtain the conditional probability that the $i$'th link in a
SAW is a rewired one, assuming that link $i-1$ is an unrewired one ($i>1$).

\begin{figure}
\vspace*{-1.7cm}
\centerline{\psfig{figure=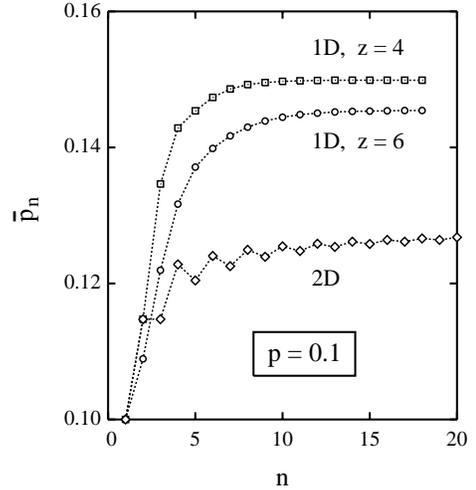,height=10.0cm}}
\vspace*{-1.5cm}
 \caption{
Conditional probability $\bar{p}_n$ that the $n$'th step in a SAW be a rewired link,
assuming that link $n-1$ is an unrewired one, as derived for $p=0.1$ from the
analytical procedure described in the text. $\bar{p}_n$ is presented as a function
of the step $n$ for small-world networks built up from 1D rings
with $z$ = 4 (squares) and $z$ = 6 (circles), as well as from a 2D square
lattice (diamonds). Dotted lines are guides to the eye.
} \label{f5} \end{figure}

We first note that in a regular lattice 
the ratio $c_i = u_i^0 / u_{i-1}^0$ obviously depends on $i$.
This ratio $c_i$ measures the average number of available links starting 
from the $(i-1)$-th site in a generic SAW on the underlying regular lattice,
and allows us to calculate the number of possible unrewired links in a
SAW on a rewired network.  
Taking into account that the fraction of unrewired connections in the whole
rewired network is $1-p$, then the mean number $q_i$ of available unrewired 
links going out from site $i-1$ in a SAW is $q_i = c_i (1-p)$.
On the other side, the mean number of rewired links going out from an
arbitrary node reached by an unrewired link is given by (see Appendix A)
\begin{equation}
w_i = p (z - \frac12) \, .
\label{wi} 
\end{equation}
(Contrary to $q_i$, the mean number $w_i$ is independent of $i$.)
Therefore, the conditional probability that link $i (> 1)$ is a rewired one, 
assuming that link $i-1$ is an unrewired one, is given by
\begin{equation} 
\bar{p}_i \equiv \frac{w_i}{w_i+q_i} = \frac {z' p} {z'p + c_i (1-p)}  \, ,
\label{barp} 
\end{equation}
with $z' = z-1/2$. For $i = 1$, we take $\bar{p}_1 = p$. 
This probability $\bar{p}_i$ is shown in Fig. \ref{f5} for the three types
of networks considered here, for a rewiring probability $p = 0.1$.

Thus, the average number of $i$-steps SAW's that do not include
rewired connections is
\begin{equation} 
A_i = (1-\bar{p}_1) \, ... \, (1-\bar{p}_i) \, u_i^0  \, ,
\label{ai} 
\end{equation}
and the number of those consisting of $i-1$ unrewired links and a rewired
one in step $i$ is
\begin{equation} 
B_i = (1-\bar{p}_1) \, ... \, (1-\bar{p}_{i-1}) \, \bar{p}_i \, u_i^0 \, .
\label{bi} 
\end{equation}
Now we note that a rewired connection in step $i$ in a (large)
small-world network ends on a random site of the network (most
probably far away from the sites already visited in the same walk).
This means that in step $i+1$ one begins most probably with a situation 
similar to that found in step $i = 1$.
Finally, the average number of $n$-steps SAW's is given by
\begin{equation}
u_n = \sum_{i_1 + ... + i_j = n} B_{i_1} B_{i_2} ... B_{i_{j-1}} A_{i_j} \, ,
\label{un2} 
\end{equation}
where the sum is extended to all possible combinations of indexes 
${i_1, ...,i_j}$ with sum equal to $n$, including null indexes, for
which we have $B_0 = A_0 = 1$.
Each term in Eq. (\ref{un2}) represents a sequence of unrewired
and rewired links, and thus the sum includes $2^n$ terms.
The SAW's corresponding to the general term in the sum include $j-1$
rewired links (in steps number $i_1, i_1+i_2, i_1+...+i_{j-1}$).
As an example, Eq. (\ref{un2}) gives for $n = 2$: 
$u_2 = B_2 + B_1^2 + B_1 A_1 + A_2$.

Note that the above equations, although rather accurate, are not exact. 
There are two reasons for this: First, each ratio $c_i$ is an average number 
of allowed links starting from a site reached in $i-1$ steps, but the actual 
number of such allowed links depends on the particular site under consideration. 
Second, finite-size effects should appear,
unless the considered networks are large enough.

Values of connective constants $\mu$ derived from $u_n$ obtained
with this procedure are presented in Figs. \ref{f3} and \ref{f4} as a function of
the rewiring probability $p$ (dashed lines).  There appears to be good 
agreement with $\mu$ derived from numerical simulations (symbols) for
$p < 0.2$. For larger $p$ the connective constants deduced from the analytical
method are larger than those yielded by the simulations. (For $z=6$ there
is a region around $p=0.3$ where the analytical results are slightly lower than
those found for the simulated networks.)

 For $p = 1$ our analytical procedure gives $\mu = z$, as for random
networks. On the other side, our numerical simulations
for small-world networks gave in all cases in the limit $p = 1$ connective
constants $\mu$ clearly lower than the mean connectivity $z$, as indicated above
and shown in Fig. \ref{f3}. This difference between both procedures is due
to the above-mentioned fact that simulated networks with $p=1$ do
not have a poissonian connectivity distribution, which is implicitly assumed
in the analytical method for this value of $p$. As a matter of fact, 
in this case we have $A_i = 0$ for $i>0$, $B_i = 0$ for $i>1$, and 
$B_1 = z$, giving in Eq. (\ref{un2}) $u_n = z^n$, as for random networks
[see Eq. (\ref{unrd})].
In this sense, our analytical procedure to calculate the number of SAW's
gives a better interpolation between regular lattices and random graphs.

In the context of this analytical approach, it is natural to expect
close to $p=0$ a linear dependence of the connective constant $\mu$ on $p$,
as found from our numerical simulations (see Sec. III), irrespective of 
the underlying lattice.
By expanding the probability $\bar{p}_i$ [Eq. (\ref{barp})] to first 
order in $p$ one finds for $i > 1$: $\bar{p}_i = z'p/c_i + O(p^2)$. 
Introducing this expression for $\bar{p}_i$ into Eqs. (\ref{ai}), 
(\ref{bi}), and (\ref{un2}), and keeping terms up to first order in $p$, 
we find:
\begin{equation}
u_n =  \sum_{i=0}^n B_i A_{n-i}   \hspace{1cm}  (p \ll 1),
\label{un3}
\end{equation}  
with the corresponding linearized expressions for $B_i$ and $A_{n-i}$.
This expression includes contributions of SAW's containing zero 
(term $A_n$, for $i=0$) and one rewired links (all other terms, $i=1, ..., n$). 
In this way, close to $p = 0$ we find for the derivative $d \mu / d p$ the
values 7.6 ($d=1,z=4$), 14.3 ($d=1,z=6$), and 6.3 ($d=2$).
Thus, it is clear that the functional dependence of $\mu$ upon $p$ for $p \ll 1$ 
does not change with the dimension of the underlying lattice.

\section{Conclusions}
 Self-avoiding walks provide us with an adequate tool to study the 
long-range characteristics of small-world networks.
For large networks, the number of SAW's increases asymptotically as 
$u_n \sim \mu^n$, provided that one considers system sizes $L \gg n$.  
For small-world networks generated from a given lattice,
the effective connectivity $\mu$ ranges from the value of the regular 
lattice to $\mu = z$  for random graphs.
For small $p$ this effective connectivity follows a linear relation
$\mu = \mu_0 + a p$, $a$ being a constant dependent on the underlying lattice. 

We have developed an analytical procedure to obtain the number of SAW's
in small-world networks. This method is based on calculating probabilities 
of finding rewired or unrewired links in the walks, and gives results in
good agreement with numerical simulations for $p \lesssim 0.2$. The results of
both methods differ for larger $p$, since they assume in practice different
connectivity distributions.  Our analytical method gives in this respect a correct 
interpolation between regular lattices and random graphs. On the contrary, 
the rewiring (simulated) process gives rise to non-poissonian connectivity 
distributions, even for a rewiring probability $p=1$, yielding in this case
networks with connective constants $\mu$ lower than the average connectivity $z$. 
                       
 Both analytical calculations and simulations similar to those presented here can 
be useful to characterize 
other kinds of networks of current interest, such as scale-free networks, whose
properties are known to depend on the asymptotic form of the connectivity
distribution for large connectivities.       \\

{\bf Acknowledgments}:
The authors benefited from useful discussions with M. A. R. de Cara. 
This work was supported by CICYT (Spain) under Contract No. BFM2000-1318. \\

\appendix
 
\section{Derivation of conditional probabilities}
Here we calculate a conditional probability related to the number of
possible connections starting from a generic node in a SAW, and necessary 
to derive Eq. (\ref{wi}) in our analytical approach in Sec. IV.
In particular, we will obtain the average number of possible rewired
links going out from a generic node $X$ (so called for definiteness),
assuming that this node was reached in a SAW through an unrewired link
in step $n$.
Here we mean by possible links all those connections that are available
for step $n+1$ in a SAW (leading to nodes not previously visited in the
same walk).

In a rewired network, the links with one end on node $X$ can be
classified for our present purpose into three types:

1) Links which were not rewired and remain as in the original lattice.
The probability distribution for the number $r$ of these connections is
given by:
\begin{equation}
P_1(r)= {z \choose r} (1-p)^{r} p^{z-r}  \, , \hspace{1cm} r = 0,... , z.
\end{equation}

2) Rewired links for which the reference node $X$ was not changed.  
Following the above notation, there are $z-r$ rewired links,
from which $s$ keep one end on site $X$. The probability distribution for 
$s$ is 
\begin{equation}
P_2(s)= {z-r \choose s} \left( \frac12 \right)^{z-r} \; , 
        \hspace{1cm} s = 0, ..., z-r.
\end{equation}

3) New (rewired) links arriving at site $X$.
 The distribution of the number $v$ of these connections is (for large 
system size $N$)
\begin{equation}
P_3(v) = \frac{1}{v!} \, (t z)^v \, e^{-z t} \; , \hspace{1cm} v \geq 0,
\end{equation}
with $t = p/2$.
                                       
Thus, given a site with $r$ unrewired connections, the number of rewired
links is $x = s + v$, where $s$ depends on $r$ and $v$ is independent of
$r$. The probability distribution for $x$ is:
\begin{equation}
Q_2^{(r)}(x) = \sum_{s=0}^{s_{max}} P_2(s)  P_3(x-s) \; ,
\label{q2}
\end{equation}  
with $s_{max} = min(z-r, x)$. Then, we have
\begin{equation}
Q_2^{(r)}(x) = \left( \frac12 \right)^{z-r}  e^{-z t}
     \sum_{s=0}^{s_{max}} {z-r \choose s} \, \frac{1}{(x-s)!} (t z)^{x-s} \; . 
\label{q2b}
\end{equation}  

Hence, the probability distribution for the number of outgoing rewired
links (assuming that the incoming link was an unrewired one) is
\begin{equation}
Q_{out}(x) = \sum_{r=1}^{z} Q_{in}(r) Q_2^{(r)}(x) \; ,
\label{qout}
\end{equation}  
where $Q_{in}(r)$ is the probability of reaching a node having $r$ unrewired 
links:
\begin{equation}
Q_{in}(r) =  \frac{r}{z} {z \choose r} (1-p)^{r-1} p^{z-r} \; .
\end{equation}
[$Q_{in}(r)$ is given, apart from a normalization constant, by the product
$r P_1(r)$.]

Finally, the mean number of outgoing rewired links [calculated with
the probability distribution $Q_{out}(x)$] is
\begin{equation}
\langle x \rangle = \sum_{x=1}^{\infty} x \, Q_{out}(x)  \, .
\label{meanx}
\end{equation}
Introducing expression (\ref{qout}) into Eq. (\ref{meanx}), and after a 
straightforward but somewhat lengthy algebra, one finds the mean value:
\begin{equation}
\langle x \rangle = p (z - \frac12)  \, .
\label{meanx2}
\end{equation}
This average value $\langle x \rangle$ is called $w_i$ in the main text,
and the last result is Eq. (\ref{wi}) there.

\end{document}